\documentclass[aps,prb,twocolumn,superscriptaddress,showpacs]{revtex4}

\usepackage{exscale}
\usepackage{t1enc}
\usepackage{graphicx}
\usepackage{amsmath,amsfonts,amssymb,amsxtra}
\usepackage{bm}
\usepackage{textcomp}

\makeatletter

\DeclareRobustCommand{\chemical}[1]{%
  {\(\m@th
   \edef\resetfontdimens{\noexpand\)%
       \fontdimen16\textfont2=\the\fontdimen16\textfont2
       \fontdimen17\textfont2=\the\fontdimen17\textfont2\relax}%
   \fontdimen16\textfont2=2.7pt \fontdimen17\textfont2=2.7pt
   \mathrm{#1}%
   \resetfontdimens}}

\newcommand{\remno}{R$_{1-x}$A$_x$MnO$_3$}
\newcommand{\lasr}[2]{\chemical{La_{#1}Sr_{#2}MnO_4}}
\newcommand{\half}{\chemical{La_{0.5}Sr_{1.5}MnO_4}}

\newcommand{\TCO}{T_\text{CO}}
\newcommand{\TN}{T_\text{N}}

\newcommand{\kk}{\bm k}
\newcommand{\Q}{\bm Q}
\newcommand{\vQ}{\bm Q}
\newcommand{\q}{\bm q}
\newcommand{\QFM}{\Q_\text{FM}}
\newcommand{\QCE}{\Q_\text{CE}}

\begin{document}

\title{ Melting of magnetic correlations  in charge-orbital ordered La$_{0.5}$Sr$_{1.5}$MnO$_4$ :
competition of ferro and antiferromagnetic states }

\author{D. Senff}
\affiliation{II.~Physikalisches Institut, Universit\"at zu K\"oln, Z\"ulpicher Str.~77, D-50937 K\"oln,
Germany}

\author{O. Schumann}
\affiliation{II.~Physikalisches Institut, Universit\"at zu K\"oln, Z\"ulpicher Str.~77, D-50937 K\"oln,
Germany}

\author{M. Benomar}
\affiliation{II.~Physikalisches Institut, Universit\"at zu K\"oln, Z\"ulpicher Str.~77, D-50937 K\"oln,
Germany}

\author{M. Kriener}
\affiliation{II.~Physikalisches Institut, Universit\"at zu K\"oln, Z\"ulpicher Str.~77, D-50937 K\"oln,
Germany} \affiliation{Department of Physics, Kyoto University, Kyoto 606-8502, Japan}

\author{T. Lorenz}
\affiliation{II.~Physikalisches Institut, Universit\"at zu K\"oln, Z\"ulpicher Str.~77, D-50937 K\"oln,
Germany}

\author{Y. Sidis}
\affiliation{Laboratoire L\'eon Brillouin, C.E.A./C.N.R.S., F-91191 Gif-sur-Yvette Cedex, France}

\author{K. Habicht}
\affiliation{Hahn-Meitner-Institut, Glienicker Str.~100, D-14109 Berlin, Germany}

\author{P. Link}
\thanks{Spektrometer PANDA, Institut f\"ur Festk\"orperphysik, TU Dresden}
\affiliation{ Forschungsneutronenquelle Heinz Maier-Leibnitz (FRM II), TU M\"unchen, Lichtenbergstr. 1,
D-85747 Garching, Germany}

\author{M. Braden}
\email{braden@ph2.uni-koeln.de}
\affiliation{II.~Physikalisches Institut, Universit\"at zu K\"oln,
Z\"ulpicher Str.~77, D-50937 K\"oln, Germany}

\date{\today}

\begin{abstract}
The magnetic correlations in the charge- and orbital-ordered manganite \half~ have been studied by
elastic and inelastic neutron scattering techniques. Out of the well-defined CE-type magnetic structure
with the corresponding magnons a competition between CE-type and ferromagnetic fluctuations develops.
Whereas ferromagnetic correlations are fully suppressed by the static CE-type order at low temperature,
elastic and inelastic CE-type correlations disappear with the melting of the charge-orbital order at
high temperature. In its charge-orbital disordered phase, \half~ exhibits a dispersion of ferromagnetic
correlations which remarkably resembles the magnon dispersion in ferromagnetically ordered metallic
perovskite manganites.

\end{abstract}


\maketitle

\section{Introduction}

Charge ordering is one of the key elements to understand colossal
magnetoresistivity (CMR) in the manganite oxides. The large drop
of the electric resistivity at the metal-insulator transition can
only partially be explained by the Zener double-exchange
mechanism.\cite{millis95a} The larger part of it seems to arise
from the competition between ferromagnetic (FM) metallic and
charge-ordered insulating states, and recent experimental and
theoretical investigations focus on electronically soft phases and
phase separation
scenarios.\cite{uehara99a,moreo99a,woodward04a,milward05a,sen07a}
The metal-insulator transition can be considered as the
stabilization of the FM metallic phase over charge-ordered
insulating states by an external parameter as e.\,g. temperature
or magnetic field.\cite{tokura00a,murakami03a}

\begin{figure}
  \includegraphics[width=0.35\textwidth]{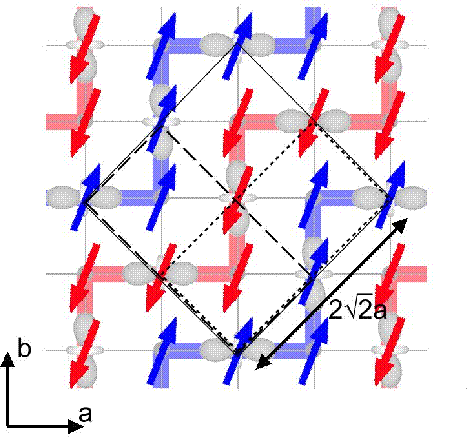}\\
  \caption{(Color online) Schematic representation of the charge, orbital and spin ordering in the CE-type
  arrangement, as proposed by Goodenough.\cite{goodenough55a} The orbital and the magnetic lattices of the
  Mn$^{3+}$-subsystem are indicated by the dotted and dashed lines, respectively. Notice that the zigzag
  chains run along the [1\,1\,0]-direction.
  }\label{Fig-Sketch-CE}
\end{figure}

At half doping, the insulating, charge-orbital ordered (COO)
phase appears most stable, and an ordered state appears as a
generic feature in the phase diagrams of cubic manganites \remno
~(R=La or a rare earth, A=Sr, Ba, Ca,...),\cite{tokura00a} as
well as in those of single and double-layered systems, as \half~
and \chemical{LaSr_2Mn_2O_7}.\cite{sternlieb96a,argyriou00a} In
spite of its relevance for the CMR effect and in spite of the
enormous number of publications in this field, the properties of
the charge-ordered state are not fully established till today.
Early investigations on La$_{0.5}$Ca$_{0.5}$MnO$_3$ by Wollan and
Koehler\cite{wollan55a} and by Goodenough\cite{goodenough55a}
proposed a checkerboard ordering of charges with sublattices of
Mn$^{3+}$ and Mn$^{4+}$ sites. Simultaneously, the single $e_g$
orbitals on the \chemical{Mn^{3+}} sites order in a stripe-like
pattern, giving rise to zigzag paths with each $e_g$ orbital
bridging two \chemical{Mn^{4+}} neighbours with $3d^3$
configuration and an empty $e_g$ level, see Fig.~
\ref{Fig-Sketch-CE}. This orbital arrangement implies a FM
ordering along the zigzag chains and an antiferromagnetic (AFM)
coupling between adjacent chains,  referred to as CE-type
ordering,\cite{wollan55a} and it explains the observed structural
and magnetic superlattice reflections in diffraction
experiments.\cite{radaelli97a,murakami98a}  More recently, an
alternative model, consisting of a coherent ordering of magnetic
dimers, called Zener-Polarons, has been proposed,\cite{daoud02a}
which is fundamentally different. Whereas in the initial model
charge and orbital ordering is located on the Mn sites, the
alternative model proposes the ordering of charges to be located
on the Mn-O-Mn bonds. The two concepts remain matter of strong
controversy
\cite{efremov04a,grenier04a,goff04a,rodriguez05a,senff06a,trokiner06a}
with the more recent work favoring the initial site-centered
model. In particular for \half~ there is strong evidence that the
bond-centered model cannot be applied.\cite{senff06a} One should,
however, consider the possibility that different manganites
exhibit different types of charge-orbital order. Furthermore, it
is important to emphasize that the site-centered model of charge
and orbital ordering is only schematic. Different
crystallographic studies
\cite{goff04a,argyriou00a,jirak85a,damay98a,jirak00a,ourselves}
find structural distortions in the charge-ordered phase which are
far smaller than what is expected for a full integer ordering
into \chemical{Mn^{3+}} and \chemical{Mn^{4+}} valencies.
Nevertheless we will use throughout this paper this
\chemical{Mn^{3+}}/\chemical{Mn^{4+}} nomenclature for clarity.

The origin of the charge-ordered state is also under debate, and
different theoretical studies focus on very different aspects. It
has been shown, that the COO state can be stabilized primarily by
cooperative Jahn-Teller distortions. In this scenario the
magnetic ordering of the CE type appears as a secondary
effect.\cite{yunoki00a,popovic02a,dong06a} On the other hand, it
has been argued, that based on anisotropic magnetic exchange
interactions the COO can be stabilized by purely electronic
effects.\cite{solovyev99a,brink99a} In this sense the COO state
is of magnetic origin, which naturally explains its melting in
magnetic fields.\cite{solovyev99a,solovyev03a,solovyev01a}

The single-layered material \half ~ is particularly well suited for the experimental investigation of
the properties of the COO state. Charge and orbital ordering occurs in this compound below
$\TCO$=220\,K and has been investigated by various
techniques.\cite{sternlieb96a,larochelle01a,mahadevan01a,wilkins03a,dhesi04a} Magnetic ordering of the
CE type occurs below $\TN$=110\,K.\cite{sternlieb96a} Compared to the perovskite manganites, the COO
state is exceptionally stable in \half~ and only very high fields of the order of 30\,{T} can melt the
ordered state implying negative magnetoresistance effects.\cite{tokunaga99a} Good metallic properties
are, however, never achieved in the single layered manganites \lasr{1-x}{1+x}, neither by magnetic
field nor by doping.\cite{moritomo95a,larochelle05a}

In a recent work we have studied the magnetic excitation spectrum
of the CE-type ordering in \half~ at low
temperatures.\cite{senff06a} The analysis of the spin-wave
dispersion is fully consistent with the classical charge and
orbital-order model\cite{goodenough55a} and underlines the
dominant character of the FM intrachain interaction: The magnetic
structure has to be regarded as a weak AFM coupling of stable FM
zigzag elements. In this article we address the thermal evolution
of the CE magnetic ground state and report on the development of
the static and dynamic magnetic correlations as studied in
neutron scattering experiments and in macroscopic measurements:
The magnons of the static CE order transform into anisotropic
short-range magnetic correlations remaining clearly observable for
$\TN<T<\TCO$. Here magnetic correlations can be described by a
loose AFM coupling of FM zigzag-chain fragments. These CE-type
fluctuations compete with isotropic ferromagnetic correlations
between $\TN$ and $\TCO$, and they fully disappear upon melting
the COO state above $\TCO$. Instead, in the charge- and
orbital-disordered phase above $\TCO$ we find purely FM
correlations, which remarkably resemble those observed in the
metallic FM phases in cubic manganites.

\section{Experimental}
\half~ crystallizes in a tetragonal structure of space-group symmetry $I4/mmm$ with room-temperature
lattice constants $a=3.86{\text{\AA}}$ and $c=12.42{\text{\AA}}$.\cite{senff05a} For most of the
neutron scattering experiments we used the same crystal as for the analysis of the spin-wave
dispersion.\cite{senff06a} The thermodynamic measurement and some parts of the neutron scattering
experiments were done with a different sample. All crystals were grown using the same floating-zone
technique as described in Ref.~ \onlinecite{reutler03a}. Elastic neutron scattering experiments were
performed at the thermal double-axis diffractometer 3T.1 and at the triple-axis spectrometer G4.3, both
installed at the Laboratoire L\'eon Brillouin (LLB) in Saclay. Selected scans measured at the high-flux
instrument 3T.1 were repeated with the same neutron energy $E$=14.7\,{meV} at the G4.3 spectrometer
with an energy resolution $\Delta E\lesssim0.6$\,{meV} to estimate the influence of slow magnetic
fluctuations. The double-axis spectrometer integrates over a sizeable energy interval, but all data
taken on both instruments agree quantitatively very well, suggesting that the diffuse magnetic
scattering is associated with time scales longer than $\sim$10$^{-11}$\,sec. Data using polarized
neutrons were acquired at the FLEX spectrometer at the Hahn-Meitner Institut (HMI) in Berlin. Inelastic
neutron data were collected on the spectrometers 1T, 2T and 4F, installed at the thermal and cold
sources at the LLB, and on the cold instrument PANDA at the Forschungsreaktor FRM II in Munich. At all
instruments the (0\,0\,2) Bragg reflection of pyrolytic graphite (PG) was used as a monochromator and
to analyze the energy of the scattered neutrons. The energy on the analyzer side was always fixed to
$E_f=14.7$\,{meV} at the thermal instruments, and typically to $E_f=4.66$\,{meV} on the cold machines.
To suppress spurious contaminations by second harmonic neutrons an appropriate filter, either PG or
cooled Beryllium, was mounted in front of the analyzer. In most of the measurements, the sample was
mounted with the tetragonal $c$ axis vertical to the scattering plane, so that scattering vectors
$(h\,k\,0)$ were accessible. Some data were collected in a different scattering plane defined by the
[1\,1\,0] and [0\,0\,1] directions of the tetragonal structure.

Specific-heat measurements were carried out using a home-build
calorimeter working with a ``continuous heating'' method.
Magnetization was measured in a commercial vibrating sample
magnetometer and electric resistivity by standard four-contact
method.

\section{Results}
Before starting the discussion of our results we illustrate the different structural and magnetic
superstructures of the COO state with the aid of Fig.~ \ref{Fig-Sketch-CE}. Below $\TCO$ the
checkerboard arrangement of the nominal \chemical{Mn^{3+}} and \chemical{Mn^{4+}} sites doubles the
structural unit cell to lattice spacings $\sqrt{2}a\times\sqrt{2}a$ with $a\sim3.8$\AA \ the lattice
constant of the original tetragonal cell. The concomitant orbital ordering reduces the symmetry and the
nuclear lattice becomes orthorhombic with lattice constants $2\sqrt{2}a$ along [1\,1\,0] and
$\sqrt{2}a$ along [1\,-1\,0]. The ordering of charges and orbitals is related to superstructure
reflections with $\kk_\text{CO}=\pm(\tfrac{1}{2}\,\tfrac{1}{2}\,0)$ and
$\kk_\text{OO}=\pm(\tfrac{1}{4}\,\tfrac{1}{4}\,0)$ in diffraction experiments, respectively. We
emphasize once more that the interpretation of the superstructure in terms of integer charge and
orbital ordering is only qualitative. The analysis of the real structural distortion reveals much
smaller effects than expected for an integer charge ordering and still needs quantitative studies and
analyzes. Considering the magnetic ordering, the CE-type structure can be divided into two sublattices
distinguishing between the two magnetic species. For the Mn$^{3+}$ ions the magnetic unit cell is of
the same size as the nuclear one, $\sqrt{2}a\times 2\sqrt{2}a$, but it is rotated by $90^\circ$ with
the long axis along [1\,-1\,0] and propagation vectors
$\kk_{\text{Mn}^{3+}}=\pm(\tfrac{1}{4}\,{-\tfrac{1}{4}}\,0)$. Therefore, the Mn$^{3+}$ spins and the
the orbital lattice contribute at different $\Q$ positions, e.\,g.~ there is a magnetic contribution at
$\Q=(0.75\,0.25\,0)\equiv (0.25\,{-0.25}\,0)$, but not at $\Q=(0.25\,0.25\,0)$, where the orbital
lattice contributes. The Mn$^{4+}$ spins contribute to neither of these positions, but to positions
indexed by $\kk_{\text{Mn}^{4+}}=\pm(\tfrac{1}{2}\,0\,0)$ and $\pm(0\, \tfrac{1}{2}\,0)$. The full
magnetic cell has to be described in a pseudocubic lattice with $2\sqrt{2}a$ along [1\,1\,0] and
[1\,-1\,0], as shown in Fig.~ \ref{Fig-Sketch-CE}. However, the orthorhombic distortion of the
tetragonal symmetry induces a twinning due to the orbital ordering in a sample crystal, as the zigzag
chains can either run along [1\,1\,0] (orientation I) or along [1\,-1\,0] (orientation II), and the
arrangement described above is superimposed by the same, but rotated by $90^\circ$. Both twin
orientations contribute equally strong in our samples, but for the analysis we will always refer to the
orientation I depicted in Fig.~ \ref{Fig-Sketch-CE}. We emphasize that the twinning due to the COO
orthorhombic distortion is the only one occuring in \half , whereas the octahedron tilt and rotation
distortions in the perovskite manganates imply a complex twinning with up to twelve superposed domain
orientations.

In a scattering experiment the superposition of both twin orientations mixes structural and magnetic
contributions at a quarter-indexed position. The magnetic contribution of orientation I is superimposed
by the orbital contribution of orientation II and vice versa. Both contributions can, however, be well
separated using polarized neutrons. In the classical polarization analysis spin-flip scattering (SF) is
always magnetic, whereas non spin-flip scattering (NSF) can be either magnetic or structural: Magnetic
moments aligned perpendicular to both, the scattering vector $\Q$ and the neutron's polarization $\bm
P$, contribute to the SF channel, those aligned parallel to $\bm P$ to the NSF channel.\cite{moon69a}
Table \ref{Table-COO-FLEX} summarizes the results of the longitudinal polarization analysis of selected
superstructure reflections at $T=5$\,{K} determined at the FLEX spectrometer for three different
choices of the neutron quantization axis, $\bm P||\Q$ $(x)$, $\bm P\bot\Q$ and within the $ab$ plane
$(y)$, and $\bm P\bot\Q$ and perpendicular to the $ab$ plane $(z)$. In addition to the quarter-indexed
reflections, Table \ref{Table-COO-FLEX} also includes the half-indexed reflection $\Q=(0.5\,1\,0)$ and
the integer-indexed reflection $\Q=(2\,0\,0)$. These reflections are entirely magnetic, respectively
nuclear, and serve as reference positions for the analysis of the quarter-indexed reflections,
providing an estimate of the experimental accuracy with flipping ratios FR=I$^\text{SF}$:I$^\text{NSF}$
of the order of 15. Inspecting the distribution of magnetic intensity at $\Q=(0.5\,1\,0)$ in the
various $\bm P_j$ channels immediately clarifies, that the magnetic moments are confined to the $ab$
planes; a canting of the moments out of the planes must be less than ${\sim}5^\circ$, in good agreement
with other estimations.\cite{sternlieb96a} Knowing the experimentally determined FR's, the scattering
observed at a quarter-indexed position can be decomposed into magnetic and structural contributions,
see the last column of Table \ref{Table-COO-FLEX}. With increasing $|\Q|$ the magnetic scattering is
suppressed following the square of the form factor and, simultaneously, the structural component is
enhanced. For small $|\Q|$, the observed intensity is, however, entirely of magnetic origin, and in the
following we may associate any scattering appearing around $\Q=(0.75\,0.25\,0)$ with magnetic
correlations.

\begin{table}
  \begin{ruledtabular}
  \begin{tabular}{cccrrc}
                 $\Q$           &     & \multicolumn{1}{c}{$P_x$} & \multicolumn{1}{c}{$P_y$}
                                & \multicolumn{1}{c}{$P_z$}
                                & $\text{I}_\text{mag}/\text{I}_\text{struc}$ \\[0.5ex] \hline
             (0.75\,0.25\,0)    &  SF & 14511 &  1087 & 14619 &            \\
                                & NSF &   796 & 14179 &   848 &  $1.00\,/\,0.00$   \\[0.5ex]
             (0.75\,0.75\,0)    &  SF &  1625 &   161 &  1624 &            \\
                                & NSF &   208 &  1659 &   264 &  $0.94\,/\,0.06$    \\[0.5ex]
             (1.25\,0.25\,0)    &  SF &  3013 &   475 &  3194 &            \\
                                & NSF &   486 &  3085 &   498 &  $0.91\,/\,0.09$    \\[0.5ex]
             (1.75\,0.25\,0)    &  SF &  8550 &  4475 &  7508 &            \\
                                & NSF & 72665 & 76331 & 72745 &  $0.05\,/\,0.95$    \\[0.5ex] \hline
             (0.5\,1\,0)        &  SF & 12999 &  1200 & 13091 &            \\
                                & NSF &   797 & 12653 &   734 &  $1.00\,/\,0.00$   \\[0.5ex]
             (2\,0\,0)          &  SF &  2486 &  1903 &  1800 &            \\
                                & NSF & 30032 & 30846 & 31304 &  $0.00\,/\,1.00$   \\[0.5ex]
  \end{tabular}
  \end{ruledtabular}
  \caption{Results of the longitudinal polarization analysis for various magnetic and structural
  Bragg positions at $T=5$\,{K}. For each $\Q$ position the observed neutron intensity is shown for the different
  choices of the neutron quantization axis $\bm P_j$, $\bm P||\Q$ (x), $\bm P\bot\Q$ and within (y), and $\bm P\bot\Q$
  and perpendicular to the scattering plane (z), and the spin flipper on (SF) or off (NSF). The last column
  gives the calculated decomposition of the observed intensity into magnetic and structural components.
  }
  \label{Table-COO-FLEX}
\end{table}

\subsection{Elastic magnetic scattering}

\begin{figure}[t]
  \centering
  \includegraphics[width=0.475\textwidth]{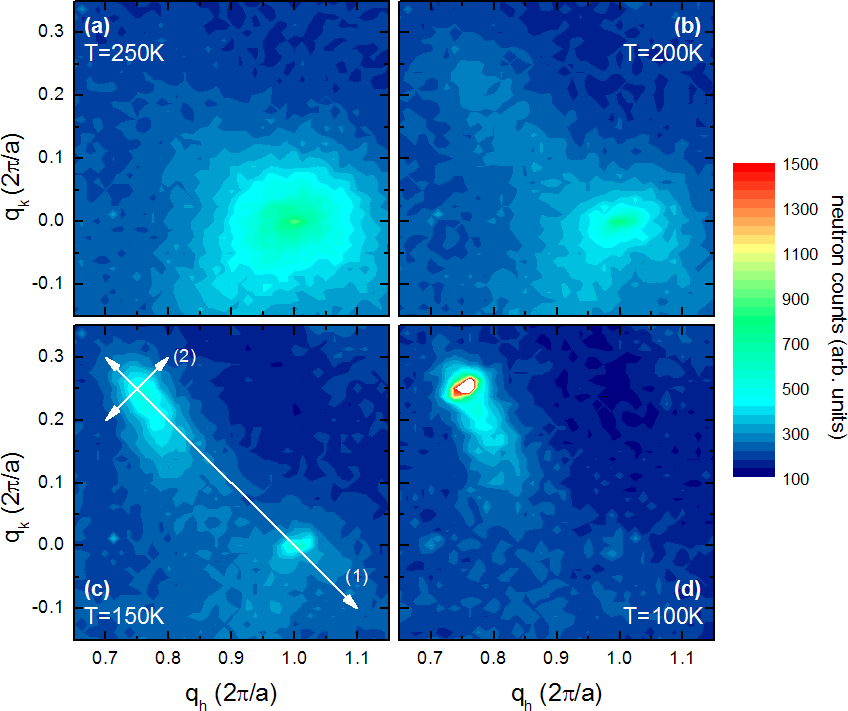}\\
  \caption{(Color online) Intensity mappings of the elastic magnetic scattering around the magnetic CE-type
  position $\Q_\text{mag}=(0.75\,0.25\,0)$ and around the two-dimensional FM zone center $(1\,0\,0)$
  at various temperatures (a) above the charge-orbital ordering at
  $T=250$\,{K}, (b,c) below $\TCO$ but above the N\'eel transition at $T=200$\,{K}, and $T=150$\,{K},
  and (d) below the AFM transition at $T=100$\,{K}. All maps were calculated from a grid of $41\times41$
  data points with $\Delta q_h=\Delta q_k=0.0125$. The two arrows in (c) denote the directions of the scans
  investigated in more detail, scan (1) senses the stacking of zigzag fragments in the CE phase and scan (2) their length.
  }\label{Fig-COO-Maps}
\end{figure}

The thermal evolution of the static magnetic correlations around $\Q=(0.75\,0.25\,0)$ has been
determined in the elastic neutron scattering experiments at the spectrometers 3T.1 and G4.3 at the LLB.
Within the estimated experimental energy resolution "static" refers to magnetic correlations on a time
scale longer than $\sim$10$^{-11}${sec.} In Fig.~ \ref{Fig-COO-Maps} we show mappings of the reciprocal
space around the magnetic CE position $\QCE=(0.75\,0.25\,0)$ including the FM position $\QFM=(1\,0\,0)$
for four different temperatures, above $\TCO$ in the charge- and orbital-disordered phase at 250\,{K},
in the COO phase above $\TN$ at 200\,{K} and at 150\,{K}, and below $\TN$ in the CE-ordered state at
100\,{K}. All four maps exhibit strong magnetic response, and the comparison of the different
temperatures directly reveals drastic changes in the character of the magnetic correlations.

In the charge- and orbital-disordered phase at $T=250$\,{K}, Fig.~ \ref{Fig-COO-Maps}a, the magnetic
scattering appears as a broad and isotropic feature centered around $\Q_\text{FM}=(1\,0\,0)$. In the
K$_2$NiF$_4$ type structure corresponding to space group $I4/mmm$, $(1\,0\,0)$ is not a
three-dimensional Bragg point due to the body-centered stacking of the MnO$_2$ layers. But when
neglecting any magnetic inter-layer coupling, i.\,e.~ analyzing magnetic correlations in a single
layer, any $(1\,0\,q_l)$ is a two-dimensional Bragg position sensing FM in-plane correlations. From the
width of the signal in the $(h\,k\,0)$ plane a nearly isotropic in-plane correlation length
$\xi_\text{iso}\approx8{\text{\AA}}$ can be estimated for ferromagnetic clusters, see below. From the
$Q_l$ dependence of the FM signal at $(0\,0\,Q_l)$ studied on a slightly under-doped sample
\cite{reutler-unpublished}, we may deduce the fully two-dimensional nature of the FM scattering and
that the moments are aligned predominantly within the planes in accordance with the anisotropy of the
magnetic susceptibility.

\begin{figure}[t]
  \centering
  \includegraphics[width=0.425\textwidth]{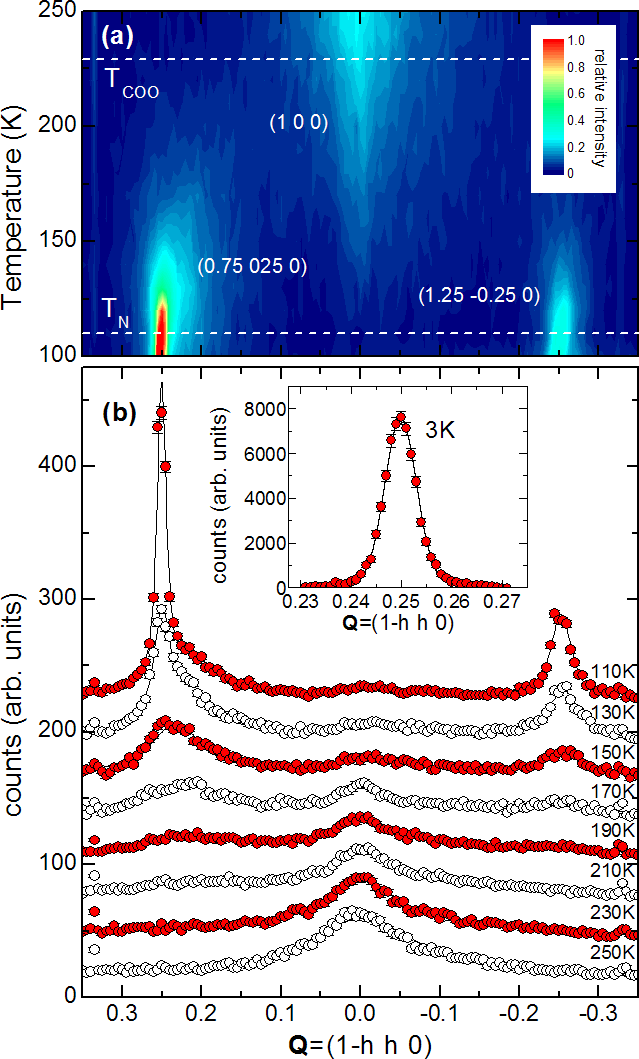}\\
  \caption{(Color online) Temperature dependence of the elastic magnetic intensity along scan 1, perpendicular to the zigzag
  chains. (a) Contour plot derived from a grid of data with $\Delta T=10$\,K and $\Delta
  q_k=0.005\tfrac{2\pi}{a}$. (b) Representative raw-data scans underlying the contour plot. For clarity,
  the data are successively shifted vertically by 300\,counts. The inset gives the profile of the magnetic
  Bragg reflection at low temperatures, T=2.5\,{K}. Lines correspond to fits as described
  in the text. In all data a minor contamination by second harmonic neutrons centered at $\Q=(1\,0\,0)$ is
  subtracted. The different scattering of the two CE-type
  reflections arises from the Mn form-factor and geometry
  conditions.
  }\label{Fig-COO-Perp}
\end{figure}

With the transition into the COO phase the magnetic correlations abruptly change. At $T=200$\,{K},
Fig.~ \ref{Fig-COO-Maps}b, the FM signal around $\QFM$ has drastically lost intensity, while,
simultaneously, magnetic intensity is increased along the path $(1\,0\,0) \rightarrow(0.75\,0.25\,0)$.
Upon further cooling, intensity is transferred from $\QFM$  to the quarter-indexed position
$\QCE=(0.75\,0.25\,0)$ and at $T=150$\,{K} two separate features are well distinguished in the mapping,
Fig.~ \ref{Fig-COO-Maps}c. The transition into the CE phase at $\TN$ finally completely suppresses the
FM response and at 100\,{K} all magnetic scattering is centered at the magnetic CE-type Bragg
reflection $\QCE$, Fig.~ \ref{Fig-COO-Maps}d.

To further analyze the competition between FM- and CE-type correlations we studied the temperature
dependence along the two lines depicted in Fig.~ \ref{Fig-COO-Maps}c in more detail. Scan 1 runs along
[1\,-1\,0] and connects $\QFM$ and $\QCE$, while scan 2 is oriented perpendicular along [1\,1\,0] and
crosses scan 1 at $\QCE$. As around $\QCE$ only the magnetic scattering of the twin orientation with
the zigzag chains running along [1\,1\,0] contributes (orientation I), scan 1 probes the magnetic
correlations perpendicular to the chains, i.\,e.~ the stacking of the FM zigzag chains within the
MnO$_2$ planes. In contrast, scan 2 determines the correlations parallel to the chains.

Let us start with the discussion of the thermal evolution of the magnetic scattering along scan 1, see
Fig.~ \ref{Fig-COO-Perp}. At the highest temperature investigated, $T=250$\,{K}, the spectrum consists
of a broad, Lorentzian-shaped feature centered at $\QFM=(1\,0\,0)$, as is already evident in Fig.~
\ref{Fig-COO-Maps}a. This signal is due to ferromagnetic planar correlations of limited correlation
length. With decreasing temperature the signal stays roughly unaffected, until charge and orbital
ordering sets in at $\TCO\approx220$\,{K}. Below $\TCO$ the signal at $\QFM$ looses spectral weight and
additional weak and very broad features become apparent around
(1$\mp\varepsilon$\,$\pm\varepsilon\,0)$. The latter features continuously sharpen, gain in intensity
and shift outward until they finally lock into the commensurate CE-type positions with
$\varepsilon=\pm0.25$ close to $\TN\approx110$\,{K}. Within the magnetically ordered phase below $\TN$,
we do not find any evidence for FM correlations anymore and the AFM CE-type reflections become sharp
and resolution limited at low temperature, Fig.~ \ref{Fig-COO-Perp}b.

\begin{figure}[b]
  \centering
  \includegraphics[width=0.495\textwidth]{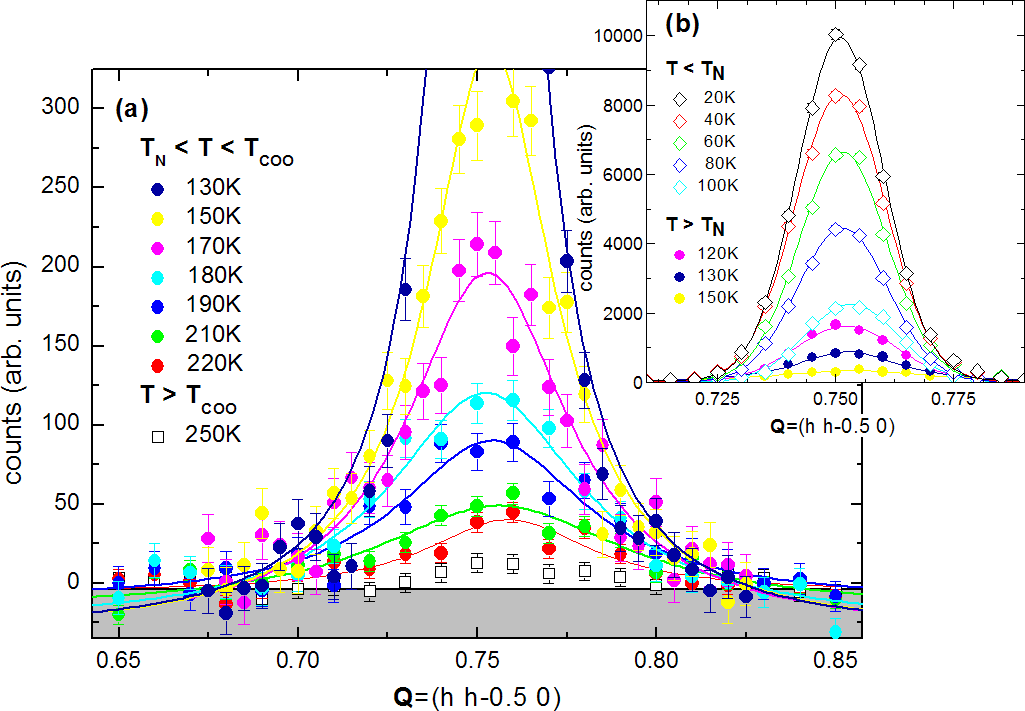}\\
  \caption{(Color online) Temperature dependence of the elastic magnetic intensity along scan 2, parallel to the zigzag
  chains, at selected temperatures (a) above and (b) below the N\'eel transition at $\TN=110$\,{K}. Note that for
  the scans at 150\,{K} and 130\,{K} the scale on the ordinate is the same in both panels. In all data a
  common background is subtracted. Lines correspond to fits with either Lorentzians or Gaussians as
   discussed in the text.}\label{Fig-COO-Parallel}
\end{figure}

The thermal evolution along scan 2, i.\,e.~ along [1\,1\,0] and parallel to the zigzag chains, is shown
in Fig.~ \ref{Fig-COO-Parallel}. At $T=250$\,{K} no signal is observable at $\QCE$. However, below the
COO transition a magnetic signal emerges, which is easily separated from the background level already
at 220\,{K}, i.\,e.~ more than 100\,K above $\TN$. Note, that there is no structural component in the
scattering at these $|\Q|$ values, see above. The magnetic signal at $\QCE$ rapidly sharpens and
increases in intensity, reflecting the transfer of spectral weight from $\QFM$ along the path
$(1\,0\,0) \rightarrow (0.75\,0.25\,0)$ to $\QCE$. The magnetic phase transition at $\TN$ is evidenced
by the change of the line shape, below 110\,{K} the profile changes from a Lorentzian into a Gaussian
with the width determined by the experimental resolution. The intensity of the reflection increases
monotonically down to the lowest temperature investigated, $T=3$\,{K}.

\begin{figure}
  \centering
  \includegraphics[width=0.40\textwidth]{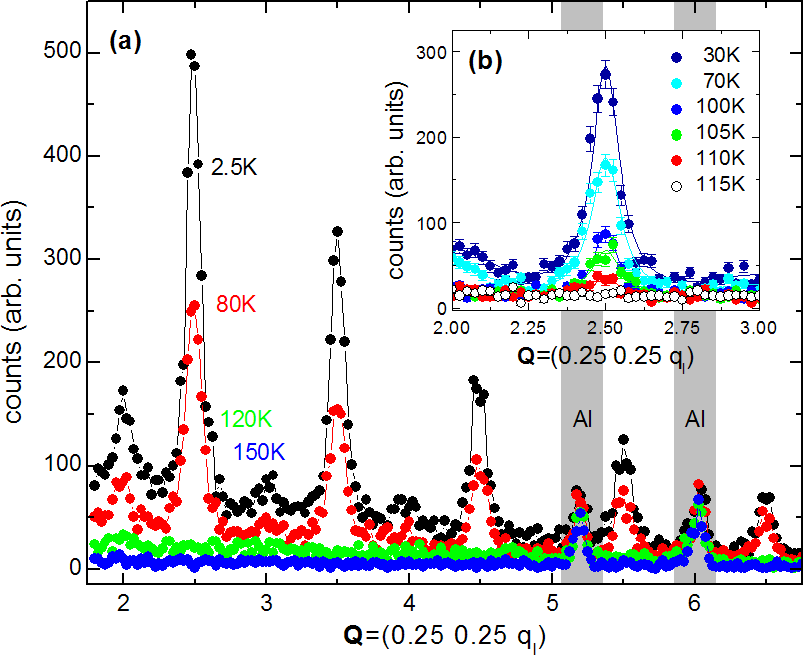}\\
  \caption{(a) (Color online) Raw-data scans along the line $(0.25\,0.25\,q_l)$ for various temperatures below $\TCO$.
  Gray-shaded areas mark spurious contributions by the scattering from Aluminium. (b) Raw-data
  scans along [0\,0\,1] centered around $q_l=2.5$ for various temperatures close to $\TN$. Lines denote
  fits with Lorentzians.
  }\label{Fig-COO-3d}
\end{figure}

So far we have only discussed the magnetic correlations within the \chemical{MnO_2} layers. In Fig.~
\ref{Fig-COO-3d} we show raw-data scans along $\Q=(0.25\,0.25\,q_l)$ for various temperatures below
$\TCO$ aiming at the magnetic correlations along [0\,0\,1]. In contrast to the in-plane correlations,
the scans along [0\,0\,1] are not structured  above $\TN$. The CE-type magnetic correlations are
entirely two-dimensional for $T>\TN$. Below the magnetic phase transition at $\TN$ a well defined
structure develops along $q_l$, and two distinct sets of reflections centered around half- and
integer-indexed $q_l$ values are detected. Both types of reflections can be associated with a different
stacking along the $c$ axis, and the observed distribution of intensity with the half-indexed $q_l$
values dominating agrees well with former observations.\cite{sternlieb96a,larochelle05a} Along
[0\,0\,1] the profile of the reflections is always significantly broader than the experimental
resolution, indicating a finite correlation length perpendicular to the layers of about 50\,\AA~ even
at lowest temperatures.

\begin{figure}
  \includegraphics[width=0.45\textwidth]{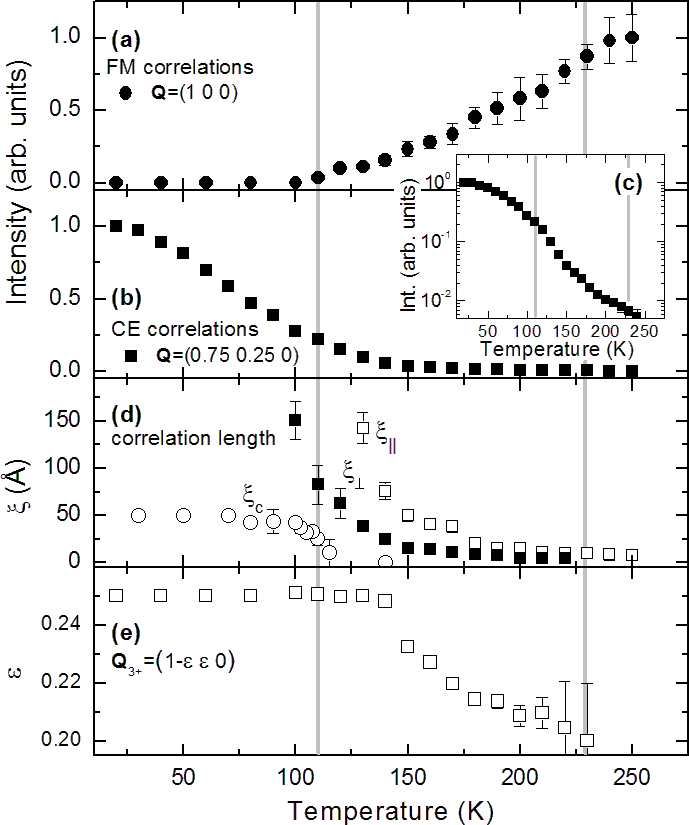}\\
  \caption{Summary of the results of the analysis of the elastic magnetic scattering showing the
  temperature dependence of (a) the intensity observed at the FM position $\Q=(1\,0\,0)$, and (b,c) at the
  AFM position $\Q=(0.75\,0.25\,0)$ on a linear, and logarithmic scale, (d) of the determined
  correlation length $\xi$ in a direction parallel to the chains, $\xi_{||}$, perpendicular to the
  chains, and within the $ab$ plane, $\xi_{\bot}$, and along the tetragonal axis, $\xi_c$, and (e) of
  the position of the AFM-signal along the line $\Q=($1$\mp\varepsilon$\,$\pm\varepsilon\,0)$.
  }\label{Fig-COO-Neutrons}
\end{figure}

For a quantitative analysis of the diffuse magnetic scattering we modeled all spectra assuming
Lorentzian or, at lower temperatures, Gaussian line shapes for the different contributions. The
integrated intensity of the reflection then directly determines the square of the magnetic order
parameter, and the width, corrected for resolution effects, is proportional to the inverse of the
magnetic correlation length. The results of this analysis are summarized in Fig.~
\ref{Fig-COO-Neutrons}. Structural superlattice reflections probing the order parameter of the orbital
and the charge ordering, which set the frame for the discussion of the magnetic correlations, appear
below $\TCO=221(1)$\,{K}, in good agreement with the
literature.\cite{larochelle01a,wilkins03a,dhesi04a} The temperature dependencies of the intensity of
the FM scattering at $\QFM=(1\,0\,0)$ and that of the CE-type reflection $\QCE=(0.75\,0.25\,0)$ as
determined from the scans presented in Fig.~ \ref{Fig-COO-Perp} and Fig.~ \ref{Fig-COO-Parallel} are
shown in Fig.~ \ref{Fig-COO-Neutrons}a\,--\,c. With the transition into the COO phase at $\TCO$ the
intensity of the FM correlations decreases and vanishes close to $\TN$. The in-plane correlation length
of the FM signal is isotropic and temperature independent, $\xi_\text{iso}\approx8{\text{\AA}}$. Below
$\TN$ no FM signal can be detected anymore in our neutron scattering experiments. In contrast, the AFM
correlations of the CE type emerge with the transition into the COO phase and compete with the FM
phases between $\TCO$ and $\TN$. Even in this pure half-doped material the transition between the
high-temperature FM correlations and the CE-type ordered state seems to occur via a microscopic phase
separation.\cite{milward05a,sen07a} We may not fully exclude that the coexistence of CE-type and FM
scattering is caused by a canting or a rotation of the spins within a single magnetic cluster, but the
different $\vQ$ shape of the scattering as well as the different temperature dependencies of the
associated correlation lengths render such an explanation unlikely.

The observed peak-height of the magnetic scattering at the CE-type Bragg reflection does not exhibit a
clear anomaly at $\TN$ and the thermal evolution appears continuous in the entire temperature regime
below 220\,{K}, although the major intensity increase is found below $\TN$. The temperature dependence
of the magnetic reflections, for example (0.5,1,0), sensing the magnetic order of the Mn$^{4+}$ sites
perfectly scales with that of the quarter-indexed ones sensing the Mn$^{3+}$ order confirming the close
coupling of the two magnetic sublattices forming the complex CE-type magnetic ordering. A clear
indication for the magnetic phase transition at $\TN$ is, however, seen in the behavior of the magnetic
correlation lengths, Fig.~ \ref{Fig-COO-Neutrons}d. The AFM correlations of the CE type exhibit a
pronounced anisotropy and temperature dependence. Both in-plane correlation lengths, $\xi_{||}$
parallel and $\xi_{\bot}$ perpendicular to the zigzag chains, rapidly increase and finally diverge as
the temperature decreases towards $\TN\approx110$\,{K}. For $T>\TN$ $\xi_{||}$ is always larger than
$\xi_{\bot}$, indicating that the intrachain correlations are much better defined than the interchain
correlations. Close to $\TN$ the ratio between $\xi_{||}$ and $\xi_{\bot}$ is most pronounced attaining
a factor of $\sim$4, and $\xi_{||}$ diverges at slightly higher temperatures than $\xi_{\bot}$. The
magnetic transition at $\TN$ has to be interpreted as the coherent AFM ordering of preformed FM zigzag
chains. The ratio of $\xi_{||}/\xi_{\bot}$ perfectly reflects the ratio of the magnetic interaction
parameters in the CE phase at low temperature, where the interaction along the chain is by far
dominating \cite{senff06a}. The out-of-plane correlation length for the CE ordering, $\xi_c$, is also
included in Fig.~ \ref{Fig-COO-Neutrons}d. In contrast to the in-plane correlations, $\xi_c$ remains
finite at lowest temperatures, $\xi_c\approx50{\text{\AA}}$, and the inter-layer correlation rapidly
disappears above $\TN$. Finally, Fig.~ \ref{Fig-COO-Neutrons}e displays the evolution of the AFM peak
position along the line (1$\mp\varepsilon$\,$\pm\varepsilon\,0)$. The incommensurability $\varepsilon$
directly reflects the AFM coupling between adjacent or more distant zigzag chains. With the onset of
the magnetic correlations near $\TCO$, $\varepsilon$ increases monotonically and locks slightly above
$\TN$ into the commensurate value $\varepsilon=0.25$. Hence, the modulation wavelength perpendicular to
the zigzag chains decreases upon cooling until at $\TN$ adjacent chains couple antiferromagnetically.
The asymmetric shape of the CE-type diffuse scattering close to $\TN$, see e.\,g.~
Fig.~\ref{Fig-COO-Perp}b, has to be ascribed to the asymmetric distribution of $\varepsilon\le0.25$:
The stacking of the zigzag chains cannot occur with a repetition scheme shorter than nearest-neighbor
chains, which is $2\sqrt{2}a$ corresponding to $\varepsilon=0.25$.

We emphasize that the well defined, three-dimensional magnetic CE-type order together with the full
suppression of the FM response is reminiscent of the high quality of our crystal. An earlier sample of
\half~ \cite{reutler-unpublished} as well as the crystal used in Ref.~\onlinecite{sternlieb96a} exhibit
significantly reduced CE-type correlation lengths. The high quality of the sample is further documented
by a clear specific-heat anomaly observed at $\TCO$, see below.

\subsection{Thermodynamic properties}

\begin{figure}[b]
  \centering
  \includegraphics[width=0.45\textwidth]{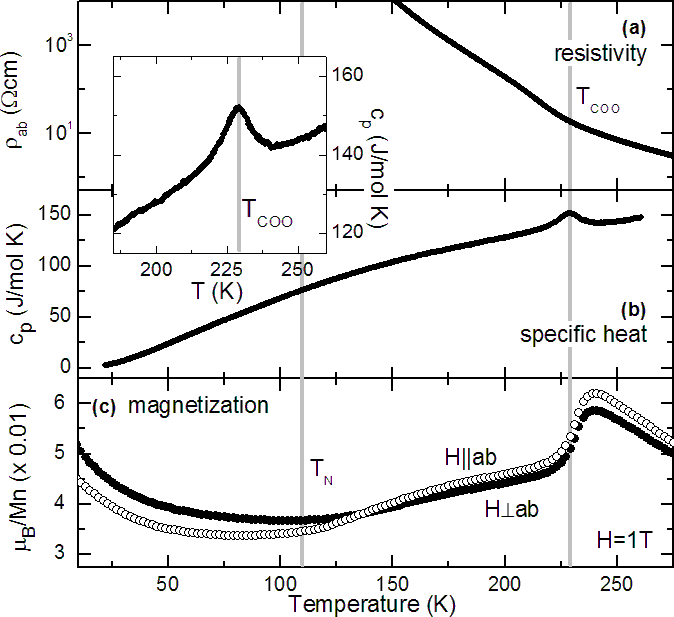}\\
  \caption{(a) Temperature dependence of the in-plane electric resistivity $\rho_{ab}$, (b) the specific
  heat $c_p$, and (c) the macroscopic magnetization for a field $H=1$\,{T} applied parallel and
  perpendicular to the $ab$ planes. Vertical grey lines mark $\TCO$ and $\TN$ as determined in the
  neutron scattering experiments.
  }\label{Fig-COO-Macro}
\end{figure}

The unusual evolution of the magnetic state with static short-range correlations appearing 100\,{K}
above $\TN$ also affects the thermodynamic quantities. In Fig.~ \ref{Fig-COO-Macro} we show the
temperature dependence of the electric resistivity $\rho_{ab}(T)$ along the planes, the specific heat
$c_p(T)$, and the macroscopic $dc$ magnetization $M_\bot(T)$ and $M_{||}(T)$ for a field $H=1$\,{T}
applied perpendicular and parallel to the \chemical{MnO_2} layers, respectively. All three quantities
show a well-defined anomaly at $\TCO$, but none around $\TN$. The electric resistivity $\rho_\text{ab}$
exhibits insulating behavior with a significant jump-like increase at $\TCO$ reflecting the real-space
ordering of the charge carriers.\cite{moritomo95a} Note, that we find no hysteresis in the temperature
dependence of the resistivity at $\TCO$. The specific heat displays a pronounced anomaly at the same
temperature documenting the well-defined character of the COO transition in our \half~ crystal. Below
$\TCO$, however, the specific heat seems to be determined by phononic contributions, and it is
difficult to detect a clear signature for an additional release of entropy around the magnetic
ordering, which, however, is consistent with the formation of short-range magnetic correlations well
above $\TN$.

The macroscopic magnetization $M(T)$, Fig.~ \ref{Fig-COO-Macro}c, is directly correlated with the
neutron scattering results presented above. For $T>\TCO$, $M(T)$ increases linearly upon cooling and
$M_\bot$ is always smaller than $M_{||}$, as there is an easy plane anisotropy parallel to the MnO$_2$
layers. The magnetization reaches a maximum slightly above $\TCO$, at $\approx240$\,{K}, and it is
strongly suppressed at the transition into the COO phase. Upon further cooling
 $M(T)$ continues to decrease down to $T=90$\,{K}
and it roughly scales with the temperature dependence of the magnetic neutron intensity observed at
$\QFM$. Remarkably, there is no well defined signature of the N\'eel transition, as $M(T)$ varies
continuously across $\TN$, both for $H\bot ab$ and $H||ab$. This behavior may be ascribed to the FM
preordered zigzag chain fragments with considerable AFM coupling already above $\TN$. At the N\'eel
transition only the stacking of the zigzag chains becomes better defined. In this sense the AFM
transition may be considered as an order disorder one.\cite{solovyev01a} The ratio between $M_\bot$ and
$M_{||}$ agrees with the expectations for an AFM order of moments aligned within the $ab$ planes due to
an easy-plane anisotropy. Below $T\approx135$\,{K} $M_\bot$ is larger than $M_{||}$, whereas the
opposite is observed in the paramagnetic phase. At low temperatures, below $T\approx50$\,{K}, both
magnetization components $M_\bot$ and $M_{||}$ exhibit a pronounced Curie-like upturn. Usually, a low
temperature upturn in the magnetization is associated with sample-dependent impurities, but in the case
of \half~ it might be a generic feature as it is observed in various studies using different sample
crystals.\cite{moritomo95a}

The most prominent feature concerns the sudden magnetization drop
at the COO transition. As pointed out by Moritomo et al., the
singular behavior at $\TCO$ can be attributed to the quenching of
the double-exchange interaction with the localization of the
$e_g$ electrons into the charge-ordered state,\cite{moritomo95a}
which is supported by ESR measurements.\cite{marumoto03a} The
neutron analysis, however, clearly shows, that the FM
correlations are not just reduced at $\TCO$, but they become
replaced by the AFM CE-type fluctuations even well above the
N\'eel transition. Due to the distinct magnetic symmetries, the
magnetic transition from a FM state to the AFM CE-type order must
be of first order allowing for phase coexistence.

\subsection{Dynamic magnetic correlations}

So far we have only considered the thermal evolution of the static magnetic correlations. The
anisotropic character of the AFM correlation and the competition between AFM and FM interactions
should, however, also significantly influence the dynamic magnetic properties. The development of the
magnetic correlations at finite energy transfer has been studied in the experiments at the
spectrometers 4F, 1T, 2T, and PANDA.

In our previous work \cite{senff06a} we established the
low-energy part of the magnetic excitations in the CE-type
ordered state which can be well described by spin-wave theory. At
low temperatures, the magnon dispersion is anisotropic with a
steep dispersion along the zigzag chains, reflecting the dominant
FM interaction along this direction.\cite{senff06a} The
pronounced magnon anisotropy can be taken as a strong indication
against the bond-centered dimer model, whereas it is naturally
described within the CE-type orbital and magnetic model.

\begin{figure}
  \centering
  \includegraphics[width=0.475\textwidth]{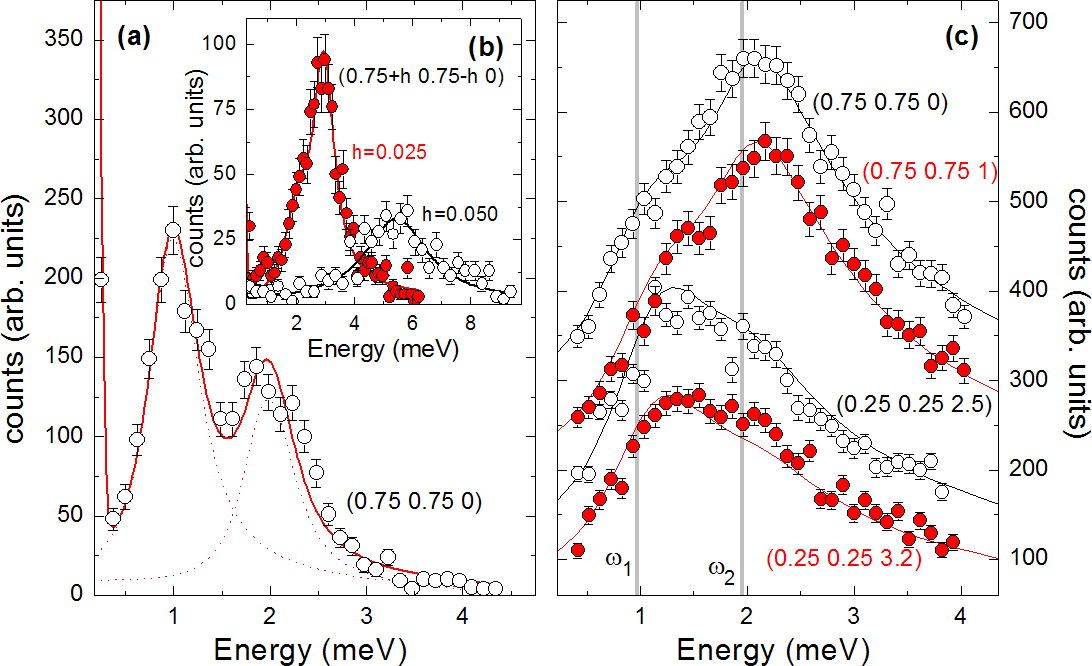}\\
  \caption{(Color online) (a) Energy scan at the AFM zone center $\Q=(0.75\,0.75\,0)$  at $T=[4]$\,{K}, and (b)
  at different positions perpendicular to the propagation of the chains at
  $\Q=(\text{0.75+$q_h$}\,\text{0.75-$q_h$}\,0)$. (c) $q_l$ dependence of the magnon signal taken with a different
  experimental setup. Lines denote fits to the data. See text for details.
  }\label{Fig-COO-Gap}
\end{figure}

In the CE-ordered phase at $T=5$\,{K} the spin-wave spectrum is gaped at the antiferromagnetic zone
center, $\q=0$, see Fig.~ \ref{Fig-COO-Gap}a, and two distinct magnon contributions can be resolved.
The degeneracy of the two AFM magnon branches seems to be removed due to  anisotropy terms. As already
seen in the high-temperature behaviour of $M(T)$, \half ~ exhibits an easy-plane single-ion anisotropy
above the COO transition. Due to the orthorhombic symmetry in the COO phase, the easy-plane anisotropy
\cite{bonesteel93a} must transform into an easy-axis symmetry, which is hidden by the twinning in the
macroscopic measurements. The magnetic anisotropy, however, is visible in the excitation spectra in the
form of the observed zone-center gap and splitting. At the antiferromagnetic zone center we find two
magnon excitations around 1.0\,{meV} and 2.0\,{meV}, Fig.~ \ref{Fig-COO-Gap}a. The splitting of the
modes is, however, restricted to the zone center, and for finite momentum $|\q|$ both modes merge
rapidly into a single excitation, Fig.~ \ref{Fig-COO-Gap}b.

To probe the character of the two zone-center modes we studied their $q_l$ dependence, Fig.~
\ref{Fig-COO-Gap}c. As only the component of the magnetization perpendicular to the scattering vector
contributes in neutron scattering, increasing $q_l$ will suppress a fluctuation polarized along the $c$
axis while a mode fluctuating perpendicular to $c$ will remain less affected or even increase in
intensity. For this purpose we used a scattering plane defined by [1\,1\,0]/[0\,0\,1], where the steep
[1\,-1\,0] branch of the dispersion points along the vertical axis. As the vertical $\bf Q$ resolution
is low on a focusing triple-axis spectrometer, this configuration integrates over a sizeable part of
the vertical dispersion strongly affecting the shape of the measured signal. In the scans taken using
this configuration, Fig.~ \ref{Fig-COO-Gap}c, the two  magnon contributions appear only as a rather
broad feature while they can be easily separated with the usual configuration with $c$ vertical to the
scattering plane. Due to the weak interlayer coupling the dispersion along $c$ is negligible and hence
the integration is very efficient with [0\,0\,1] vertical. To take the different experimental
conditions into account and to resolve the two different magnon contributions in all scans, we
convoluted the four-dimensional resolution function with the dispersion surface as derived in Ref.~
\onlinecite{senff06a} using the ResLib-code.\cite{reslib06a} In a first step we evaluated the data
taken with [0\,0\,1] vertical and refined the magnon energies at $\q=0$ yielding
$\omega_1=0.97(2)$\,{meV} and $\omega_2=1.97(4)$\,{meV}, respectively. Using these values as a starting
point we modeled the data taken in the second setup. As a first result, all data with different $q_l$
values can simultaneously be described using the same magnon frequencies. There is no measurable
spin-wave dispersion vertical to the \chemical{MnO_2} layers. As a second fit parameter we modeled the
intensity distribution at the various $q_l$ positions. With increasing $|\Q|$, the signal $\omega_1$
follows the square of the magnetic formfactor, while the signal $\omega_2$ is additionally suppressed.
Assuming the mode $\omega_2$ to be entirely polarized along $c$, we find a good agreement with the
data, and the splitting of the magnon frequencies can be fully ascribed to different magnetic
anisotropies: Fluctuations within the \chemical{MnO_2} layers are more favorable than those along the
$c$ axis as it is expected. The sizeable gap associated with the in-plane anisotropy,
$\omega_1$=0.97meV is remarkable as it documents that magnetic moments are also pinned through the
orbital anisotropy of the zigzag chains.

\begin{figure}[t]
  \centering
  \includegraphics[width=0.495\textwidth]{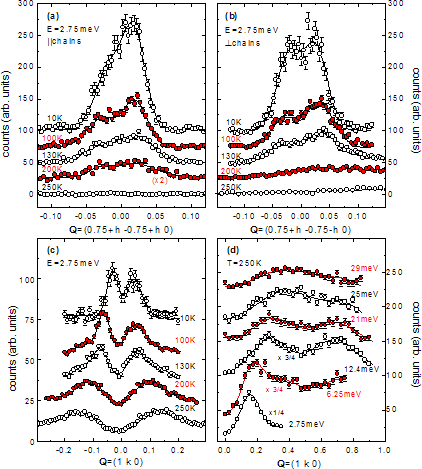}\\
  \caption{(Color online) (upper panel) $\Q$ scans at a finite energy $E=2.75$\,{meV} for different
  temperatures across the CE position $({0.75}\, {-0.75}\, 0)$ in a direction (a) parallel to the zigzag
  chains, [1\,1\,0], and (b) perpendicular to the chains along [1\,-1\,0]. (lower panel) Raw-data
  scans tracking (c) the temperature dependence of the magnetic fluctuations around the FM position
  $\QFM=(1\,0\,0)$ with $E=2.75$\,{meV}, and (d) the $\q$ dependence of the FM fluctuations for
  $T=250$\,{K}. All scans are corrected for the different Bose contributions after the substraction
  of a linear background. For clarity, subsequent scans are shifted by a constant amount on the abscissa.
  Lines correspond to fits with Gaussians.
  }\label{Fig-COO-Fluc}
\end{figure}

To determine the temperature dependence of the magnetic fluctuations we scanned the excitations around
$\QCE=(0.75\, {-0.75}\, 0)$ and $\QFM=(1\,0\,0)$ in different directions at a constant energy of
$E=2.75$\,{meV}. These scans were performed at 10\,{K} and 100\,{K} in the CE ordered phase, in the COO
phase above $\TN$ at [130]\,{K} and at 200\,{K}, and in the disordered phase at 250\,{K}, see Fig.~
\ref{Fig-COO-Fluc}. First, we discuss the thermal evolution around $\QCE$. At $T=10$\,{K} and along
[1\,1\,0], i.\,e.~ parallel to the zigzag chains, the spectrum can be decomposed into the two magnon
contributions centered at $\QCE +\q$ and $\QCE -\q$, see Fig.~ \ref{Fig-COO-Fluc}a. However, as the
dispersion in this direction is steep, both signals strongly overlap in agreement with our previous
study.\cite{senff06a} With increasing temperature, the magnon signal is suppressed following roughly
the magnetic order parameter and shifts outward due to the overall softening of the magnetic
dispersion. In consequence, the magnon contributions are fully resolved at $T=100$\,K, which is
corroborated by further scans at 4\,{meV} exhibiting a similar behavior (data not shown). Upon heating
across $\TN$ the inelastic response broadens, but we do not observe a significant change in the magnon
frequencies. At 200\,{K} the spectrum can be described by two contributions centered at the same
positions as at 100\,{K}, indicative of the stable FM interaction within the zigzag chain fragments. In
the perpendicular direction, Fig.~ \ref{Fig-COO-Fluc}b, the two magnon contributions are separated
already at 10\,{K}, as in this direction the spin-wave velocity is significantly
reduced.\cite{senff06a} Upon heating, the signal shifts outward, too, but this shift is more pronounced
than that parallel to the chains. Furthermore, for $T>\TN$ the inelastic intensity is fully smeared
out, and at $T=200$\,{K} we do not find any correlations which can be associated with the inter-zigzag
coupling, whereas, in the direction along the chains the inelastic intensity is still well centered
around $\QCE$. The inelastic CE-type magnetic correlations are thus turning one-dimensional in
character between $\TN$ and $\TCO$: Only the magnetic coupling within a zigzag fragment remains finite
close to $\TCO$. Above the charge-orbital ordering the magnetic fluctuations reminiscent of the CE-type
magnetic order are completely suppressed.

Around the FM $\Q$ point we find a fundamentally different behavior of the magnetic fluctuations.
$\QFM=(1\,0\,0)$ is also a Bragg position of the CE-type magnetic structure, and therefore inelastic
neutron scattering may detect the CE-type spin-wave modes also around (1\,0\,0) but with a strongly
reduced structure factor, see Fig.~3 in Ref.~\onlinecite{senff06a}. However, the scattering around
(1\,0\,0) does not evolve like the CE-order parameter and the associated fluctuations described above.
The inelastic intensity close to $(1\,0\,0)$ remains well-defined over the entire temperature range up
to highest temperatures, see Fig.~\ref{Fig-COO-Fluc}c. At $T=250$\,{K} the differences between the
spectra around the different $\Q$ positions are most evident: Around $\QCE$ no inelastic signal can be
detected anymore, whereas the dynamic correlations around $\QFM$ are clearly structured. These
fluctuations are entirely FM in character -- there is no evidence for CE-type correlations left at this
temperature -- and they have to be associated with the isotropic FM clusters revealed in the diffuse
magnetic scattering. As around $\QFM$ both types of magnetic correlations may contribute, CE-type as
well as FM ones, the thermal progression of the dynamics around this position, Fig.~\ref{Fig-COO-Fluc},
documents how the isotropic FM correlations compete with and finally are replaced by the CE-type
ordering (upon cooling), as it is fully consistent with the coexistence of different magnetic phases in
the elastic scattering.

\begin{figure}[b]
  \centering
  \includegraphics[width=0.425\textwidth]{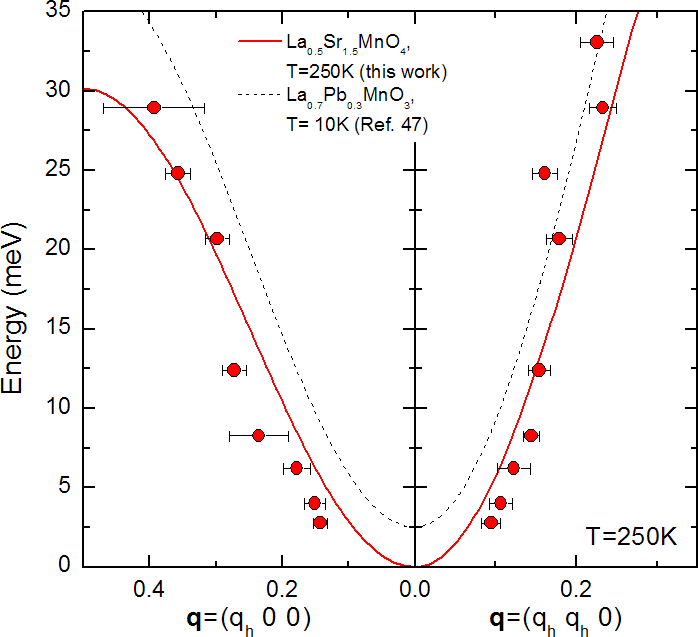}\\
  \caption{(Color online) $\q$ dependence of the FM fluctuations in the disordered phase at T=250\,{K} for
  $\q=(q_h\,0\,0)$ and $\q=(q_h\,q_h\,0)$. Solid lines denote a fit to the data using an isotropic dispersion relation
  according to Eq.~ \ref{EqDispFM}, dotted lines give the magnon dispersion in the FM metallic state (T=[10]\,{K}) of
  the perovskite \chemical{La_{0.7}Pb_{0.3}MnO_3} taken from Ref.~ \onlinecite{perring96a}.
  }\label{Fig-COO-DispFM}
\end{figure}

At T=250\,{K}, i.e. above any magnetic and charge ordering, we followed the $\q$ dependence of the FM
fluctuations up to a maximal energy of 35\,{meV} along the main symmetry directions in the MnO$_2$
layers, see Fig.~\ref{Fig-COO-Fluc}d. The inelastic response is always rather broad in $\Q$ space, and
above 12.5\,{meV} two different features overlap strongly, which are, however, centered at equivalent
$\q$ positions in neighboring FM Brillouin zones, (1\,0\,0) and (1\,1\,0). With increasing energy both
signals disperse towards the FM zone boundary, which they finally reach close to 30\,{meV} at
(1\,0.5\,0). In the diagonal direction along [1\,1\,0] the energies of the FM fluctuation extend to
even higher energies (raw data not shown), and the resulting dispersion of the FM fluctuations in the
disordered phase above the COO transition is summarized in Fig.~ \ref{Fig-COO-DispFM}.

On a square lattice, the spin-wave dispersion for a Heisenberg ferromagnet with isotropic exchange
$J_\text{iso}$ between nearest neighbors is given by:
\begin{equation}\label{EqDispFM}
    \hbar\omega(\q)= 4J_\text{iso} S(2-\cos(2\pi q_h)-\cos(2\pi q_k)).
\end{equation}
The observed $\q$ dependence of the magnetic correlations is reasonably well described by this simple
Hamiltonian with only a single nearest-neighbor interaction, underlining the isotropic character of the
FM correlations above $\TCO$. The slight overestimation of the frequencies at the low-$|\q|$ limit may
arise from the finite size of the FM clusters. We do not find evidence for an excitation gap, and the
best fit to the data yields an exchange energy $2J_\text{iso}S=7.5(5)$\,{meV}. The strength of the
ferromagnetic exchange in the disordered phase is significantly reduced compared to the FM interaction
in the CE structure, $2J_\text{FM}S\approx18$\,{meV} along the zigzag chains,\cite{senff06a} but still
points to a sizable hopping mediated through the Zener double exchange even though the single-layer
compound remains insulating. Furthermore, $2J_\text{iso}$ is well comparable with the magnetic exchange
interaction in the FM metallic phases of perovskite manganites: The spin-wave dispersion in the FM
phase of the CMR-compound \chemical{La_{0.7}Pb_{0.3}MnO_3} is included in Fig.~\ref{Fig-COO-DispFM},
which is also described by a single nearest-neighbor exchange interaction.\cite{perring96a} The
dispersion of magnetic correlations in paramagnetic \half~ and the magnon dispersion in FM
\chemical{La_{0.7}Pb_{0.3}MnO_3} are remarkably similar. Therefore, the strength of the Zener exchange
in the disordered phase above the COO transition in the insulating compound \half~ must be of similar
magnitude as that in the metallic state of the perovskite CMR compounds.\cite{ye06a} The fact that the
dispersion is perfectly isotropic parallel to the planes fully excludes the interpretation that the FM
clusters are formed by the coupling of FM zigzag chain fragments.

\section{Comparative Discussion of the temperature dependencies of the magnetic correlations}

\begin{figure}
  \centering
  \includegraphics[width=0.475\textwidth]{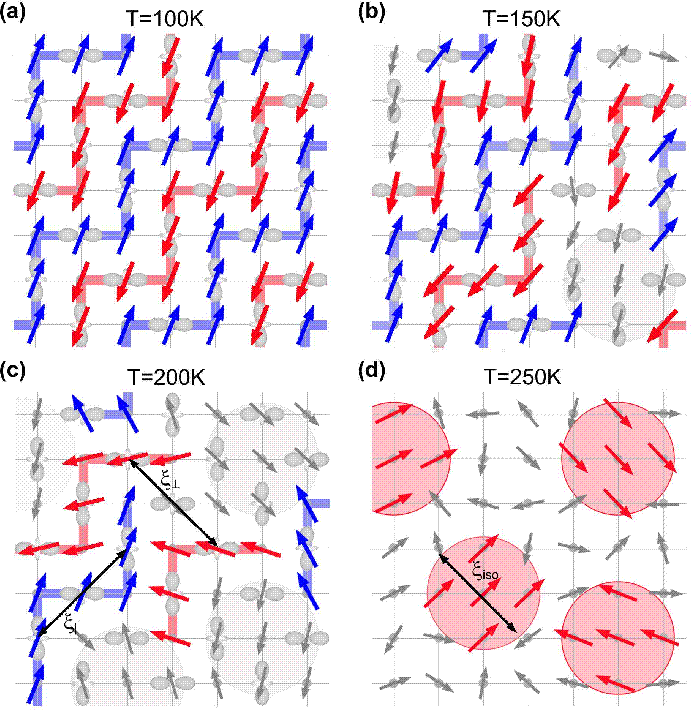}\\
  \caption{(Color online) Real-space sketches of the magnetic correlations in the \chemical{MnO_2}-layers for various
  temperatures, (a) in the long-range ordered phase below $\TN$, (b,c) in the paramagnetic, but orbital ordered
  state $\TN<T<\TCO$, and (d) in the disordered regime above $\TCO$. Qualitatively, the four
  sketches are correlated to the intensity mappings presented in Fig.~ \ref{Fig-COO-Maps}, as
  indicated by the temperatures associated with each sketch.
  }\label{Fig-COO-Sketches}
\end{figure}

The combination of the macroscopic and of the neutron scattering
studies results in a comprehensive description of the evolution
of the magnetic correlations in \half ~ upon heating across the
magnetic and the charge/orbital transition temperatures. The
untwinned nature of the \half -sample crystal -- besides the
twinning directly introduced through the COO order -- is of great
advantage, as we may easily interpret the distinct signals.
Real-space sketches of the magnetic correlations for different
temperatures are presented in Fig.~ \ref{Fig-COO-Sketches} to
illustrate the different stages of the magnetic ordering between
the short-range FM clusters at high temperature and the
well-defined CE-type magnetic order at low temperature. In the
following we will discuss the evolution of the magnetic order
with increasing temperature. Although the details of the magnetic
ordering might depend on the precise composition of the half-doped
manganite,\cite{mathieu06a} and in particular on its
single-layer, double-layer or perovskite structure, the general
aspects, how magnetic correlations evolve with temperature,
should be qualitatively the same, as previous less comprehensive
studies
indicate.\cite{bouloux81a,kumar97a,bao97a,millange00a,tomioka02a,ye05a}

At low temperature, \half ~ exhibits the well established CE-type magnetic order \cite{sternlieb96a}
with no trace of another coexisting magnetic phase, see Fig.~\ref{Fig-COO-Sketches}a. Moreover, the
magnon excitations in this phase can be well described within spin-wave theory and perfectly agree with
the site-centered charge-orbital ordering model. In \half ~ the spin-wave velocity is highly
anisotropic as the magnetic interactions along the zigzag chains are by far the
strongest.\cite{senff06a} The localized electron on the Mn$^{3+}$ site seems to yield a strong magnetic
bridging within the zigzag chains, as discussed by Solovyev.\cite{solovyev03a,solovyev01a}

The magnetic transition at $\TN$ is uncommon, as one does not find any signature of it in the
temperature dependence of the magnetization in \half ~ as well as in many half-doped perovskite
manganites.\cite{bouloux81a,kumar97a,bao97a,millange00a,tomioka02a,ye05a} The smooth variation of the
susceptibility across $\TN$ relates to the temperature dependence of the magnetic diffuse scattering.
Short-range magnetic correlations of the CE type persist above $\TN=110$\,{K} and can be observed up to
$\TCO=220$\,{K}. These correlations are purely two dimensional and restricted to single
\chemical{MnO_2} layers. Moreover, the rod-like structure of the magnetic scattering within the
$a^*b^*$ planes documents the strong  anisotropy of these two-dimensional correlations: The coupling
parallel to the chains is considerably better defined than that in the perpendicular direction in
perfect agreement with the magnetic interaction parameters deduced from the spin-wave dipersion
\cite{senff06a}. For $\TN<T<\TCO$ zigzag fragments are formed as isolated objects, see Fig.~
\ref{Fig-COO-Sketches}b and c. With the reduction of temperature down towards $\TN$, the typical length
$\xi_{||}$ of the zigzag fragments grows and interchain correlations begin to develop on a length scale
$\xi_\bot$, thereby adapting more and more Mn spins into the CE-type correlated matrix, Fig.~
\ref{Fig-COO-Sketches}c. All zigzag elements finally order at $\TN$ and the long-range CE-type ordering
establishes in the \chemical{MnO_2} layers, Fig.~ \ref{Fig-COO-Sketches}a. Perfect three-dimensional
ordering is, however, not achieved and the correlation length perpendicular to the planes remains
finite.

The magnetic transition at $\TN$ has thus to be considered as the
coherent ordering of  preformed FM zigzag-chain fragments into
the two-dimensional CE-type structure. A recent neutron
scattering study on the related material,
\chemical{Pr_{0.55}(Ca_{0.8}Sr_{0.2})_{0.45}MnO_3}, has revealed
similar features in the short-range magnetic correlations. Ye et
al.~ report on anisotropic magnetic correlations above $\TN$,
which are interpreted in terms of an electronically smecticlike
phase.\cite{ye05a} An unusual thermal evolution of the magnetic
order across $\TN$ seems to be a characteristic feature of the
CE-type ordering.

The picture of preformed FM zigzag fragments is furthermore supported by the thermal evolution of the
inelastic magnetic fluctuations, which above $\TN$ exhibit similar anisotropies as the static diffuse
scattering. Furthermore, the softening of the magnon frequencies parallel to the chains is less
pronounced than that perpendicular to the chains, once more demonstrating the predominance of the FM
coupling along the chains within the COO state.\cite{senff06a}

In between the magnetic transition at $\TN$ and charge-orbital
ordering at $\TCO$ we find a coexistence of the CE-type and FM
elastic correlations documenting, how ferromagnetism and the
CE-type order compete at these intermediate temperatures. With
increasing temperature and approaching $\TCO$ the FM correlations
get more and more weight, as it is indicated in the sketches in
Fig.~ \ref{Fig-COO-Sketches}b and c. It is remarkable that even
in this pure half-doped material with an apparently very stable
charge-orbital order, FM and CE-type elastic correlations coexist
over a wide temperature interval.  This behavior points to
microscopic phase separation.\cite{uehara99a,moreo99a,sen07a}
Although we may not fully exclude it, the interpretation of the
coexisting FM and CE-type magnetic correlation by a spin rotation
within a single cluster seems very unlikely, as the correlation
lengths behave fully differently with temperature.

In the disordered phase above $\TCO$ the CE-type elastic and
inelastic magnetic correlations are fully suppressed. This
behavior strongly contrasts with observations in cuprates or
nickelates where magnetic correlations reminiscent of stripe
order persist far into the charge disordered
phases.\cite{stripe-ni1,stripe-ni2,stripe-ni3,stripe-cu-1,cheong91a,nakajima95a,tranquada97a,bourges03a}
One has to conclude that at the charge- and orbital-order
transition at $\TCO$ not only the long-range order disappears,
but that also the dynamic COO fluctuations become suppressed. This
might be the consequence of the first order character of the
charge- and orbital-order transition in \half, which is already
imposed by symmetry.  Although our results clearly indicate that
the dynamic fluctuations of the CE-type-associated charge-orbital
order depicted in Fig.~\ref{Fig-Sketch-CE} are lost above $\TCO$,
we think that some charge and orbital ordering persists on a
local scale as it is frequently labeled through polaronic
effects. Such local electron-lattice coupling should be
associated with the strong phonon renormalization observed,
e.\,g.~ in La$_{0.7}$Sr$_{0.3}$MnO$_3$.\cite{reichardt99a}

In the charge- and orbital-disordered phase, there is, however, strong diffuse ferromagnetic scattering
proving the existence of isotropic FM clusters with an average size of 8\,${\text{\AA}}$, see Fig.~
\ref{Fig-COO-Sketches}d. The loosely antiferromagnetically bound FM zigzag chain fragments seem to
transform into these clusters upon heating across $\TCO$. This picture perfectly agrees with the jump
in the static susceptibility at $\TCO$. The magnetization of \half~ is typical for the CE-type ordering
and can be compared to those of other charge-ordered perovskite manganites with a narrow one-electron
bandwidth.\cite{kumar97a,millange00a,tomioka02a}
In addition to the suppression of the double-exchange interaction due to the charge
ordering, the onset of the AFM correlations explains the large drop of $M(T)$ at $\TCO$. The further
reduction of $M(T)$ below $\TCO$ scales with the decrease of the FM intensity in the diffraction
experiments and underlines the competition of FM and CE-type correlations.

The observed sequence of magnetic phases -- isotropic short-range ferromagnetic correlations above
$\TCO$, anisotropic correlations for $\TN<T<\TCO$ and long range ordering below $\TN$ -- is also
stabilized in various theoretical approaches.\cite{solovyev03a,brey05a,daghofer06a} Based on
anisotropic magnetic exchange interactions, Solovyev predicts a magnetically disordered state
consisting of stable FM-ordered zigzag elements for $\TN<T<\TCO$,\cite{solovyev03a} just as observed in
our neutron data. However, whether or not these magnetic fluctuations are strong enough to stabilize
the COO state, as originally proposed in Ref.~ \onlinecite{solovyev03a}, can not be finally answered
from the present data. However, we emphasize that the CE-type magnetic fluctuations (both elastic and
inelastic ones) are very weak slightly below the charge-orbital order shedding some doubt on the
interpretation that the magnetic mechanism alone is strong enough to stabilize the charge- and
orbital-ordered state.

\section{Conclusions}

In summary, we have studied the magnetic correlations in the charge- and orbital-ordered manganite
\half~ by elastic and inelastic neutron scattering techniques and by macroscopic thermodynamic
measurements.  The twin-free real structure of \half~ single crystals allows for a very precise
analysis of the diffuse scattering and of the inelastic fluctuations.

The N\'eel transition in \half ~ has to be considered as an
order-disorder transition of FM zigzag chain fragments, thereby
explaining the absence of well-defined anomalies in the
magnetization at $\TN$, which is characteristic for numerous
charge- and orbital-ordered perovskite materials as well. The
in-plane magnetic correlations in \half ~ are highly anisotropic
in character, as the magnetic coupling within the zigzag chains is
strongly dominating, whereas adjacent chains are only loosely
coupled. FM correlations are fully suppressed below $\TN$ in the
CE-type magnetic state in our high-quality crystal of \half , but
in between $\TN$ and $\TCO$ AFM CE-type and FM correlations
compete, with the FM fluctuations gaining more weight upon
approaching the charge-orbital order transition at $\TCO$. The
coexistence of diffuse FM and CE-type correlations with different
$\vQ$ shape  indicates that microscopic phase separation occurs
even in this pure half-doped material.

Inelastic fluctuations reminiscent of the CE-type order can be
followed up to $\TCO$, but only the dispersion along the zigzag
chains remains steep emphasizing once more the dominant role of
the FM interaction along the zigzag chains. With the transition
into the charge-orbital disordered phase above $\TCO$, the
CE-type elastic and the inelastic correlations become fully
suppressed. This behavior is fundamentally different from that of
the magnetic stripe-type fluctuations in the layered cuprates or
nickelates, where inelastic magnetic correlations reminiscent of
a static stripe phase can be observed far above the charge
ordering or even in samples which actually do not exhibit static
stripe ordering at all.
\cite{stripe-ni1,stripe-ni2,stripe-ni3,stripe-cu-1,cheong91a,nakajima95a,tranquada97a,bourges03a}
In \half ~ the COO transition is of first order and apparently
suppresses all traces of the complex low-temperature CE-type
magnetic ground state. Instead of the CE-type magnetic
fluctuations, strong isotropic in-plane FM correlations govern the
charge-orbital disordered phase above $\TCO$. The sizeable
elastic diffuse scattering is directly related with the large
magnetic susceptibility. In addition there are well-defined
inelastic fluctuations. It is remarkable, that the dispersion of
these FM correlations in the disordered phase of \half~ so closely
resembles the magnon dispersion of the FM metallic perovskite
phases. All our observations together underline the competition
between FM and AFM CE-type magnetic correlations in \half . These
FM and AFM states are less different in character than one might
naively think. The extra $e_g$ electron per two Mn sites in \half~
constitutes the isotropic FM interaction in the charge-orbital
disordered phase above $\TCO$; the same electron apparently also
provides the dominant oriented FM interaction when it localizes
in the orbital-ordered phase.

The COO transition at $\TCO$ is clearly associated with the
crossover between FM and AFM CE-type correlations. However it is
difficult to decide whether these different magnetic fluctuations
are just the consequence or the cause of the charge- and
orbital-order transition. The fact that CE-type correlations are
found immediately below $\TCO$ may be taken as evidence for a
magnetic mechanism of the COO transition. However, close to
$\TCO$ the CE-type fluctuations are very weak (much weaker than
at low temperature) and sizeable, even stronger FM correlations
remain closely below $\TCO$. This suggests that the transition
into the charge- and orbital-ordered state is further driven by
some non-magnetic mechanism, as e.\,g.~ by Jahn-Teller
distortions.

{\it Acknowledgments} This work was supported by the Deutsche Forschungsgemeinschaft through the
Sonderforschungsbereich 608. We thank P.~Reutler and D.~Khomskii for numerous stimulating discussions,
as well as P.~Baroni for  technical support.


\end{document}